\documentclass[runningheads]{llncs}
\usepackage[T1]{fontenc}
\usepackage{graphicx}
\usepackage{booktabs}
\usepackage[misc]{ifsym}
\newcommand{\corr}{(\Letter)}

\usepackage{amsmath,amsfonts,bm}

\def\eqref#1{equation~\ref{#1}}

\def\1{\bm{1}}

\DeclareMathAlphabet{\mathsfit}{\encodingdefault}{\sfdefault}{m}{sl}
\SetMathAlphabet{\mathsfit}{bold}{\encodingdefault}{\sfdefault}{bx}{n}

\usepackage{hyperref}
\usepackage{url}

\usepackage{caption}
\usepackage{subcaption}
\usepackage{graphicx}

\usepackage{multirow}
\usepackage{booktabs}
\usepackage{subcaption}
\usepackage{nicematrix,tikz,makecell}
\usepackage{wrapfig}
\usepackage{amsmath}
\usepackage{amsfonts}
\usepackage{bbm}
\usepackage{bm}
\usepackage{mathtools}

\usepackage{bbding} 

\usepackage{xspace}
\newcommand{\QB}{QuickBooks\xspace}
\newcommand{\our}{\texttt{Rel-Cat}\xspace}
\newcommand{\bert}{\texttt{Txn-Bert}\xspace}
\newcommand{\topknn}{\texttt{TopK NN}\xspace}

\newcommand{\code}{\emph{Code}\xspace}
\newcommand{\cat}{\emph{Category}\xspace}

\usepackage{color, colortbl}
\definecolor{Gray}{gray}{0.9}

\usepackage{xcolor}        
\definecolor{txn_test}{HTML}{B5739D}
\definecolor{txn_historical}{HTML}{7EA6E0}

\begin{document}

\title{Transaction Categorization with \\Relational Deep Learning in \QB}


\author{Kaiwen Dong\inst{1,2} \and
Padmaja Jonnalagedda\inst{1} \and
Xiang Gao\inst{1} \and
Ayan Acharya\inst{1} \and
Maria Kissa\inst{1} \and
Mauricio Flores\inst{1} \and
Nitesh V. Chawla\inst{2} \and
Kamalika Das\inst{1} \corr
}

\authorrunning{K. Dong et al.}

\institute{Intuit, Mountain View CA 94043, USA 
\email{\{kaiwen\_dong,saisri\_jonnalagedda,Xiang\_Gao,maria\_kissa,\\
mauricio\_flores,Kamalika\_Das\}@intuit.com}
\and
University of Notre Dame, Notre Dame IN 46556, USA 
\email{\{kdong2,nchawla\}mus@nd.edu}}

\toctitle{Transaction Categorization with Relational Deep Learning  in \QB}
\tocauthor{Kaiwen~Dong,Padmaja~Jonnalagedda,Xiang~Gao,Ayan~Acharya,Maria~Kissa,Mauricio~Flores,Nitesh~V.~Chawla,Kamalika~Das}

\maketitle              


\begin{abstract}
Automatic transaction categorization is crucial for enhancing the customer experience in \QB by providing accurate accounting and bookkeeping. The distinct challenges in this domain stem from the unique formatting of transaction descriptions, the wide variety of transaction categories, and the vast scale of the data involved. Furthermore, organizing transaction data in a relational database creates difficulties in developing a unified model that covers the entire database. In this work, we develop a novel graph-based model, named \our, which is built directly over the relational database. We introduce a new formulation of transaction categorization as a link prediction task within this graph structure. By integrating techniques from natural language processing and graph machine learning, our model not only outperforms the existing production model in \QB but also scales effectively to a growing customer base with a simpler, more effective architecture without compromising on accuracy. This design also helps tackle a key challenge of the cold start problem by adapting to minimal data. 

\keywords{Transaction Categorization  \and Relational Deep Learning \and Financial Applications.}
\end{abstract}


\section{Introduction}


\QB offers essential bookkeeping and accounting capabilities tailored to the needs of small and medium-sized businesses. It enables them to efficiently manage critical aspects of their business operations, including accounting, payroll, payments, and inventory. A key feature of \QB is its ability to categorize financial activities captured through invoice descriptions, bank statements, etc., flexibly, enhancing insights into business performance and streamlining tax compliance. By automating labor-intensive and error-prone tasks, \QB allows business owners to focus on driving growth and increasing revenue.

At the core of \QB's functionality is its advanced bookkeeping experience. Most business transactions today are processed through financial institutions, and \QB integrates seamlessly with these institutions, enabling businesses to link their accounts and synchronize data. This connectivity triggers an influx of transactions—approximately $6.2$ billion annually into \QB. Having business owners or accountants manually review transactions would be an ineffective use of their time. Automating the processing of such a vast volume of transactions is crucial. 

\begin{wrapfigure}{r}{0.5\textwidth}
    \vspace{-4mm}
    \includegraphics[width=0.9\linewidth]{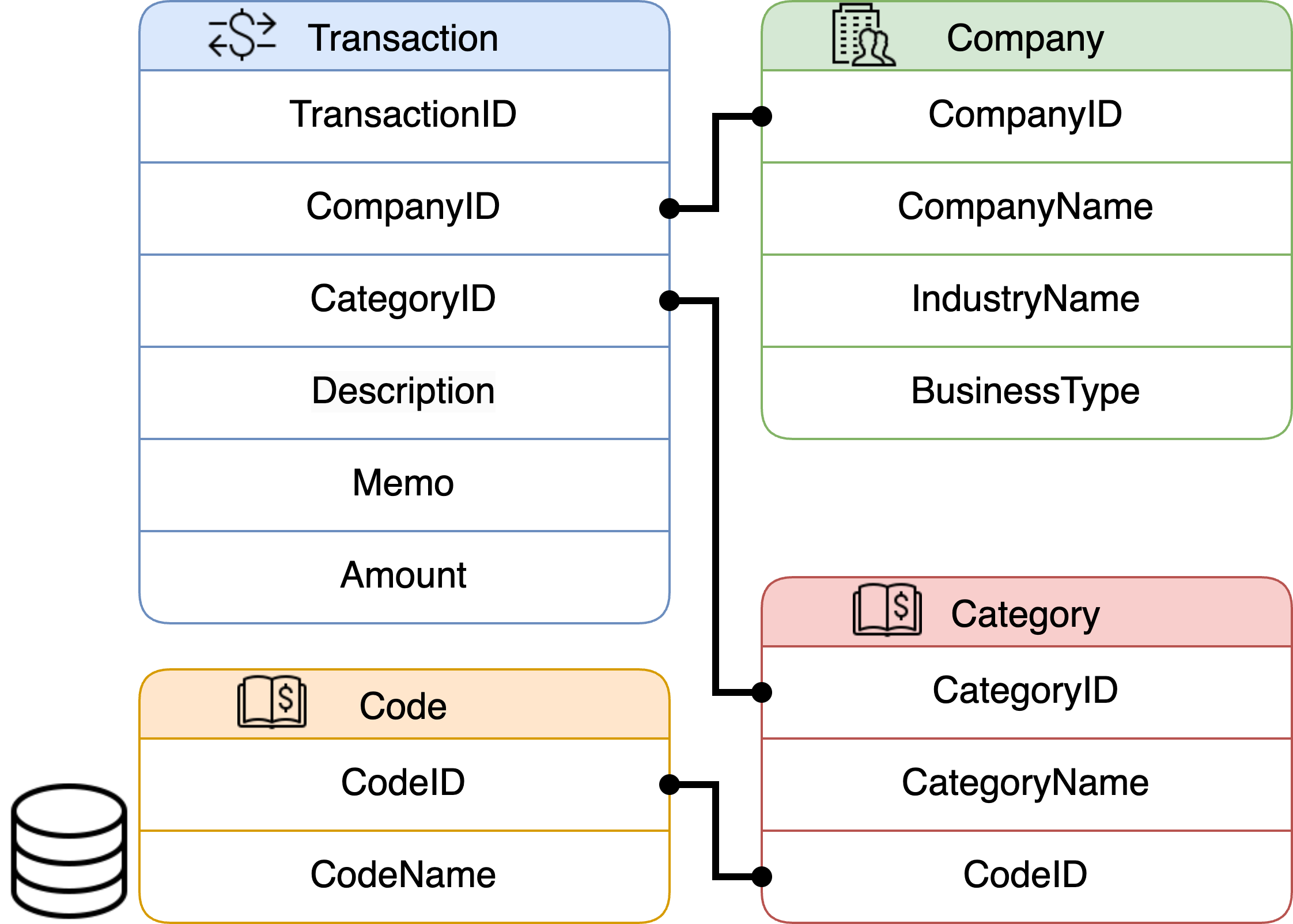}
    \caption{The schema of the relational database of \QB transactions.}
    \label{fig:schema}
    \vspace{-8mm}
\end{wrapfigure}
To facilitate its sophisticated accounting features, \QB organizes transactions into specific categories or accounts. For example, a fuel purchase at an ExxonMobil station might be categorized under ``Transportation'', while an electricity bill could be classified as ``Utilities''. This paper addresses \QB's transaction categorization challenge, employing state-of-the-art methods in natural language processing~\cite{devlin_bert_2019} and graph machine learning~\cite{bronstein_geometric_2017} to solve this as a link prediction problem~\cite{dong_pure_2023} within a relational database. Drawing inspiration from Relational Deep Learning~\cite{fey_position_2024}, we propose a unified approach to effectively model the transaction database through interconnected relational tables, introducing modifications specific to the unique challenges and practical requirements of \QB.
\subsection{Problem statement}
In \QB, effective bookkeeping relies on the accurate categorization of transactions into specific accounts. Businesses have unique needs and preferences for how transactions are categorized. \QB facilitates this by allowing customization of account names. This feature enables different companies to maintain both common and distinct account names based on their individual requirements. To further refine the organization of financial data and support compliance with tax regulations, \QB allows users to classify these account names into a structured hierarchy of more abstract account types. This system not only personalizes the accounting experience for each business but also guarantees the accurate recording of financial activities. {The more abstract account type, referred to as {\code}, corresponds to the IRS tax code, while the more granular account name, {\cat}, is user-defined. For example, \cat ``Airfare'' and ``Internet'' can both be grouped into a more abstract \code ``Business Expenses''.}


The primary task we address in this paper is predicting the appropriate \cat for any new transaction imported into \QB. In addition to delivering the most likely categorization for a new transaction, we explore providing the Top-5 probable Categories. This approach is predicated on the likelihood that the company's preferred \cat is more often found within the Top-5 predictions rather than solely the top prediction. Offering a selection of probable \cat can significantly enhance user trust in \QB's capabilities, fostering a stronger reliance on \QB for critical financial management tasks.

The data supporting this transaction categorization task is managed within a relational database, where various tables are interconnected through primary and foreign keys. The database schema, detailed in~\autoref{fig:schema}, includes the following critical tables:
\begin{enumerate}
    \setlength{\leftmargin}{0pt}
    \item \textbf{Transaction table}: Stores records of all transactions across different companies.
    \item \textbf{\cat table}: Contains all specific account names used by \QB users.
    \item \textbf{\code table}: Includes the abstract account types aligned with overarching tax codes, that facilitate the organization of~\cat.
    \item \textbf{Company table}: Contains information about companies utilizing \QB.
\end{enumerate}
\subsection{Related Works}
Transaction categorization is fundamental to the user experience in \QB. There are several ways to solve and productionize transaction categorization in accounting systems. To date, two major approaches have been deployed to effectively solve the task in \QB.

The first approach, known as IRIS~\cite{lesner_large_2019}, categorizes incoming transactions based solely on a company's historical data. It begins by extracting business entity names from transaction descriptions using a rule-based normalization process that includes case folding and digit folding. IRIS then queries the company's historical transactions for similar business entities, defining similarity via Jaccard similarity—entities are considered similar if they are frequently categorized together. A weighted voting mechanism is then employed to determine the most likely \cat for the new transaction. While IRIS is an efficient system capable of managing large datasets and ensuring quick database queries, its reliance on a company's historical data limits its utility, especially for new users with little to no transaction history.

The second methodology is embodied in the 
work by~\cite{liu_categorization_2021} in \QB.
This work addresses the limitations of IRIS by enhancing performance for users new to the system (cold-start users) with a populational model. It utilizes a Word2Vec-based~\cite{mikolov_distributed_2013} encoding for transactions and \cat, and applies a contrastive learning framework to maximize the matching pairs of transaction-\cat and minimize the non-matching pairs. 
It also employs a logistic regression classifier to enable personalized categorization for each company. During inference, the calibrated population and personalized model is applied to predict the category of a new transaction. Although this method has demonstrated effective performance in practice, maintaining an individual logistic regression classifier for each company introduces significant overhead and risks of overfitting. Additionally, calibrating two models can lead to suboptimal performance due to challenges in effectively leveraging strengths of both models.
\subsection{Challenges and Contributions}
\paragraph{Challenges.}The transaction categorization task in \QB presents several core challenges. Firstly, transaction data is stored in a relational database composed of multiple tables, making it difficult to effectively apply a single unified machine learning model without explicitly engineering features to consolidate the tables. Secondly, the transaction description field, crucial for identifying the semantic meaning of a transaction, often follows the formatting standards of financial institutions rather than natural language. This necessitates an effective method to encode transaction data, capturing nuances such as business entity names and financial acronyms. Thirdly, the vast scale and skewed distribution of \cat labels result in a highly imbalanced learning scenario. The categorization is highly personalized so that one transaction can be put into different \cat by different users. Traditional multi-class classifiers can struggle with effectiveness and generalizability on this task. Last but not the least, the large volume of transactions processed by \QB requires a balance between model capability and computational efficiency to ensure real-time performance. 
\paragraph{Contributions.} The design choices of the proposed pipeline are geared towards tackling one or more of these challenges. The salient contributions are summarized as follows:
\begin{enumerate}
    \item To model the relational database of the transaction data, we transform the database to a heterogeneous graph and apply a unified graph model, \our, to effectively represent the relationships within the data and support both new and old users of \QB. 
    \item To encode the transaction data with unique linguistic characteristics, we integrate a trained-from-scratch text encoder \bert into \our, which effectively captures the semantics of the transaction data.
    \item To handle large-scale transaction data efficiently, we introduce practical techniques to improve \our's scalability, including a similarity-based neighbor sampling, edge direction dropping, and Top-K Nearest Neighbor early exit.
\end{enumerate}






\section{\bert: Text Encoder trained from scratch}\label{sec:txn-bert}
In this section, we introduce \bert, a text encoder developed specifically for encoding transaction descriptions into fixed-length embeddings. The entire encoder is trained from scratch, recognizing the unique linguistic patterns found in transaction data. We begin with the rationale behind pretraining a new language model, followed by the detailed process of the training of the transformer model. Furthermore, we discuss the custom tokenizer training and its empirical advantages in Appendix~\ref{app:tokenizer}.

\subsection{Unique Linguistic Characteristics of Transaction Data}
To effectively convert transaction text data into embeddings, it is essential to first examine the unique characteristics of this data.

Transaction descriptions, found in banking statements, serve as crucial identifiers for financial activities within businesses. These descriptions adhere to formatting standards set by financial institutions. For instance, Visa mandates that business entity names in transaction descriptions be no longer than 25 characters, requiring abbreviations as necessary.\footnote{https://usa.visa.com/content/dam/VCOM/download/merchants/visa-merchant-data-standards-manual.pdf}.

Such descriptions, therefore, consist predominantly of abbreviated business names, a feature markedly distinct from the more fluid and expansive natural language. This distinctiveness necessitates a specialized approach for encoding transaction. The conventional language models, typically pretrained on generic natural language datasets, do not suffice for capturing the nuances of transaction descriptions. This inadequacy forms the core motivation for developing \bert from the ground up, ensuring it is finely tuned to the specific lexicon and syntax of transaction language.


\subsection{Training the transformer model}
\begin{wrapfigure}{r}{0.5\textwidth}
    \vspace{-8mm}
    \includegraphics[width=1.0\linewidth]{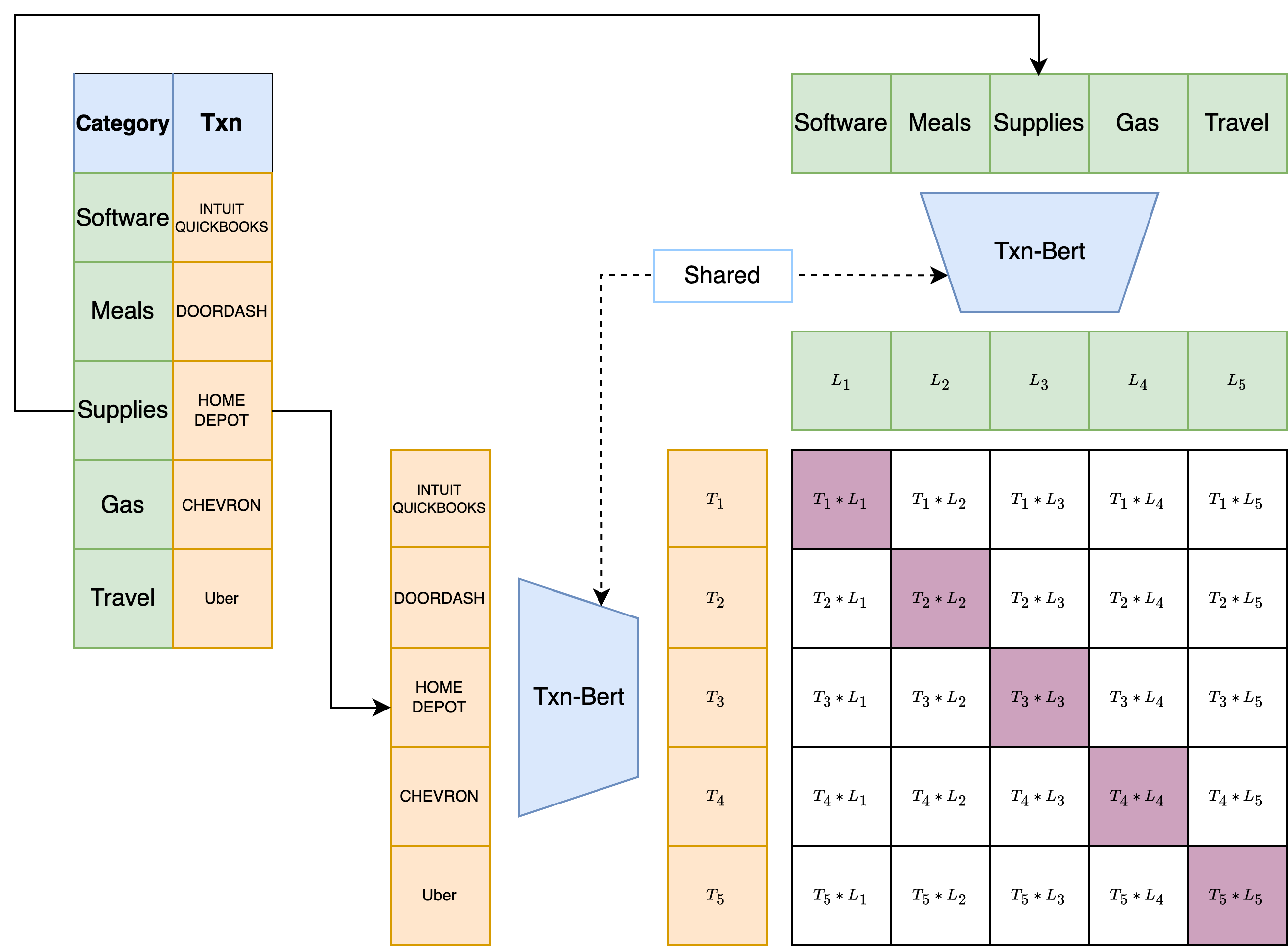}
    \caption{Training paradigm of \bert.}
    \label{fig:bert_training}
    \vspace{-6mm}
\end{wrapfigure}
We start by building a custom tokenizer to fit the specific vocabulary of transaction descriptions. We refer the details of building such tokenizer to Appendix~\ref{app:tokenizer}. Once we have built the custom tokenizer, the next step is to train our text encoder. We train the model to classify new transactions into specific \cat solely based on the information from that transaction—like its description, amount, and any additional memo—without considering the historical patterns of the company. We model our training approach after Sentence-Bert~\cite{reimers_sentence-bert_2019}, treating it as a sentence pair matching problem to match transactions to \cat. The training paradigm of \bert is shown in~\autoref{fig:bert_training}. We adopt a Siamese network architecture, where a shared text encoder independently processes a pair of a transaction and its corresponding \cat.

We format each transaction by combining its key fields into a single sentence. We include the <description>, <amount>, and <memo>. We also add a <polarity> field, which labels the transaction as ``received'' if the amount is positive or ``paid'' if it's negative. The complete transaction text looks something like this: ``Transaction <polarity> \$<amount> for: <description> <memo>.'' For the \cat labels, we use them just as they are.


The text encoder is a transformer~\cite{vaswani_attention_2017}, which will refine the representation of each token iteratively through the transformer blocks. After processing through these layers, we take a mean pooling strategy to get a single representation for the whole sentence. The configuration of the transformer can be found in Appendix~\ref{app:bert_config}.

The training objective mirrors the CLIP approach~\cite{radford_learning_2021}, optimizing for high cosine similarity between matched transaction-\cat pairs and low similarity for unmatched pairs. We use a symmetric cross-entropy loss. Specifically, given a batch of $N$ transaction-\cat pairs $\{(T_i,L_i)|1\leq i \leq N\}$, and our text encoder $f(\cdot)$, the training loss is:
\begin{align*}
    \mathcal{L} = -\frac{1}{N}\sum_{i=1}^{N}& \log \left( \frac{e^{\text{Sim}{(f(T_i),f(L_i))}}}{\sum_{j=1}^{N} e^{\text{Sim}{(f(T_i),f(L_j))}}} \right) + \log \left( \frac{e^{\text{Sim}{(f(T_i),f(L_i))}}}{\sum_{j=1}^{N} e^{\text{Sim}{(f(T_j),f(L_i))}}} \right)
\end{align*}
where $\text{Sim}{(v,w)} = \frac{v \cdot w}{|v| \cdot |w|}$ represents the cosine similarity.This method leverages all non-matching pairs within a batch for contrastive learning, thus eliminating the need for explicit negative sampling. This effective training allows \bert to robustly encode transaction and \cat data for use in downstream tasks.


\section{\our: Modeling relations within database}
\begin{figure*}[t]
    \centering
    \resizebox{1.0\linewidth}{!}{%
    \includegraphics[width=0.82\linewidth]{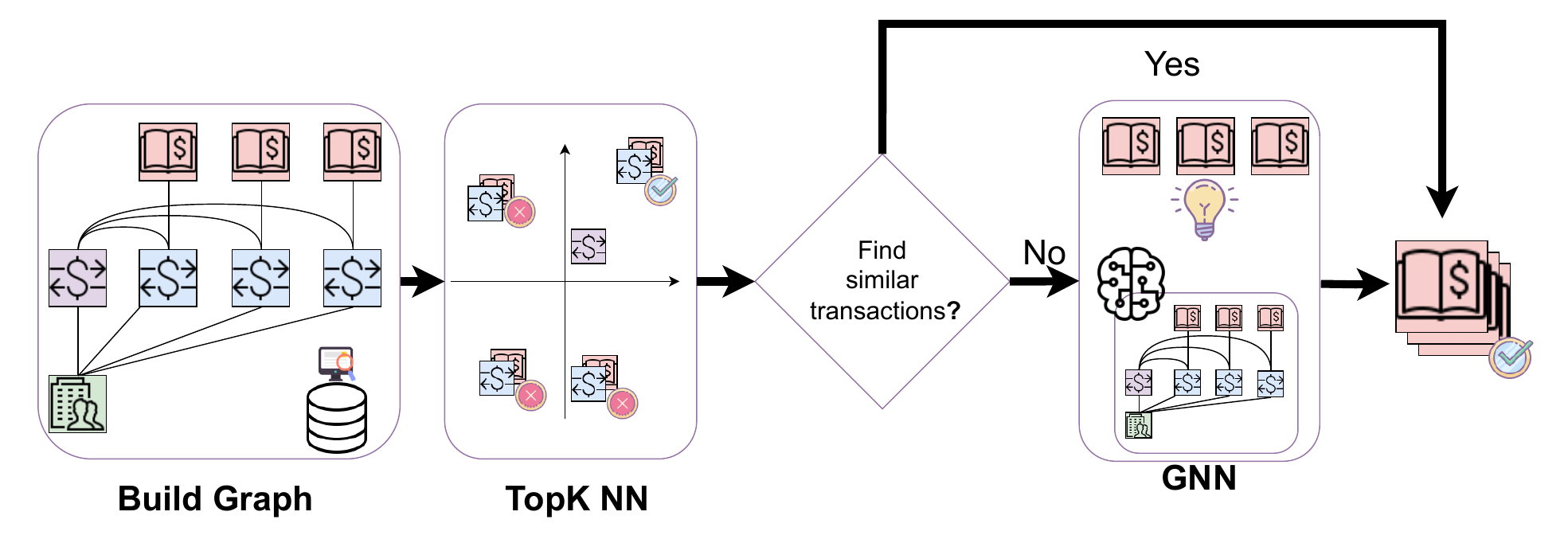}
    }
    \caption{The overview pipeline of \our.}
    \label{fig:framework}
    \vspace{-6mm}
\end{figure*}

In this section, we introduce~\our. We outline our approach for preprocessing a relational database into a heterogeneous graph, thereby redefining the transaction categorization task as a link prediction problem within this graph. We then describe~\our, a hybrid method combining rule-based early exit (\topknn) with a robust heterogeneous graph neural network (GNN). The overview pipeline can be found in~\autoref{fig:framework}.

\subsection{Build a heterogeneous graph from a relational database}\label{sec:build_graph}
The transaction data in \QB is organized within a relational database comprising multiple interconnected tables, each representing different facets of the transaction data. To unify modeling across this multi-table architecture, we employ the concept of Relational Deep Learning~\cite{fey_position_2024}, transforming the relational database into a heterogeneous graph. We provide an overview of this transformation process with a conversion diagram depicted in~\autoref{fig:convert}.

\subsubsection{Conversion overview}
Referencing~\autoref{fig:schema}, the relational database for transaction data consists of four key components:
\begin{enumerate}
    \item \textbf{Multiple tables}: The database is structured into several tables, each detailing a specific aspect of the transactions.
    \item \textbf{Rows within tables}: Each table comprises multiple rows, each row encapsulating a distinct transaction or fact.
    \item \textbf{Foreign-primary key relationships}: Rows across different tables are interconnected via foreign-primary key associations, facilitating relational references among them.
    \item \textbf{Row attributes}: Each row includes attributes that describe its elements, such as the transaction description in the transaction table or the company name in the company table.
\end{enumerate}

These elements of the relational database are mapped to the components of a heterogeneous graph as follows:

\begin{enumerate}
    \item \textbf{Node types}: Each table in the database is treated as a distinct node type within the graph.
    \item \textbf{Nodes}: Individual rows within a table are represented as nodes with the node type corresponding to their table.
    \item \textbf{Edges}: Connections between rows (nodes) across tables, facilitated by foreign-primary key pairs, are represented as edges in the graph.
    \item \textbf{Node attributes}: Attributes of each row, particularly textual data, are utilized as node attributes in the graph.
\end{enumerate}
Following the method described in~\cite{fey_position_2024}, we convert a relational database $(\mathcal{T}, \mathcal{L})$ into a heterogeneous graph $G=(\mathcal{V}, \mathcal{E}, \phi, \psi)$. The relational database consists of tables $\mathcal{T}=\{T_i\}$, each containing rows that become graph nodes $v \in \mathcal{V}$, with node types assigned by table origin via $\phi$. Attributes $x_v$ of each row $v$ are encoded using \bert embeddings. Edges $\mathcal{E}$ between nodes represent primary-foreign key relationships across tables defined by links $\mathcal{L}$, with relation types $\mathcal{R}=\mathcal{L}\cup\mathcal{L}^{-1}$ accounting for both original and inverse relations via $\psi$. Full details of the conversion procedure are provided in the Appendix~\ref{app:convert}.

\subsubsection{Transform to a link prediction task}
\begin{figure*}[t]
    \centering
    
\begin{subfigure}[t]{0.38 \linewidth}
    \centering
    \includegraphics[width=\linewidth]{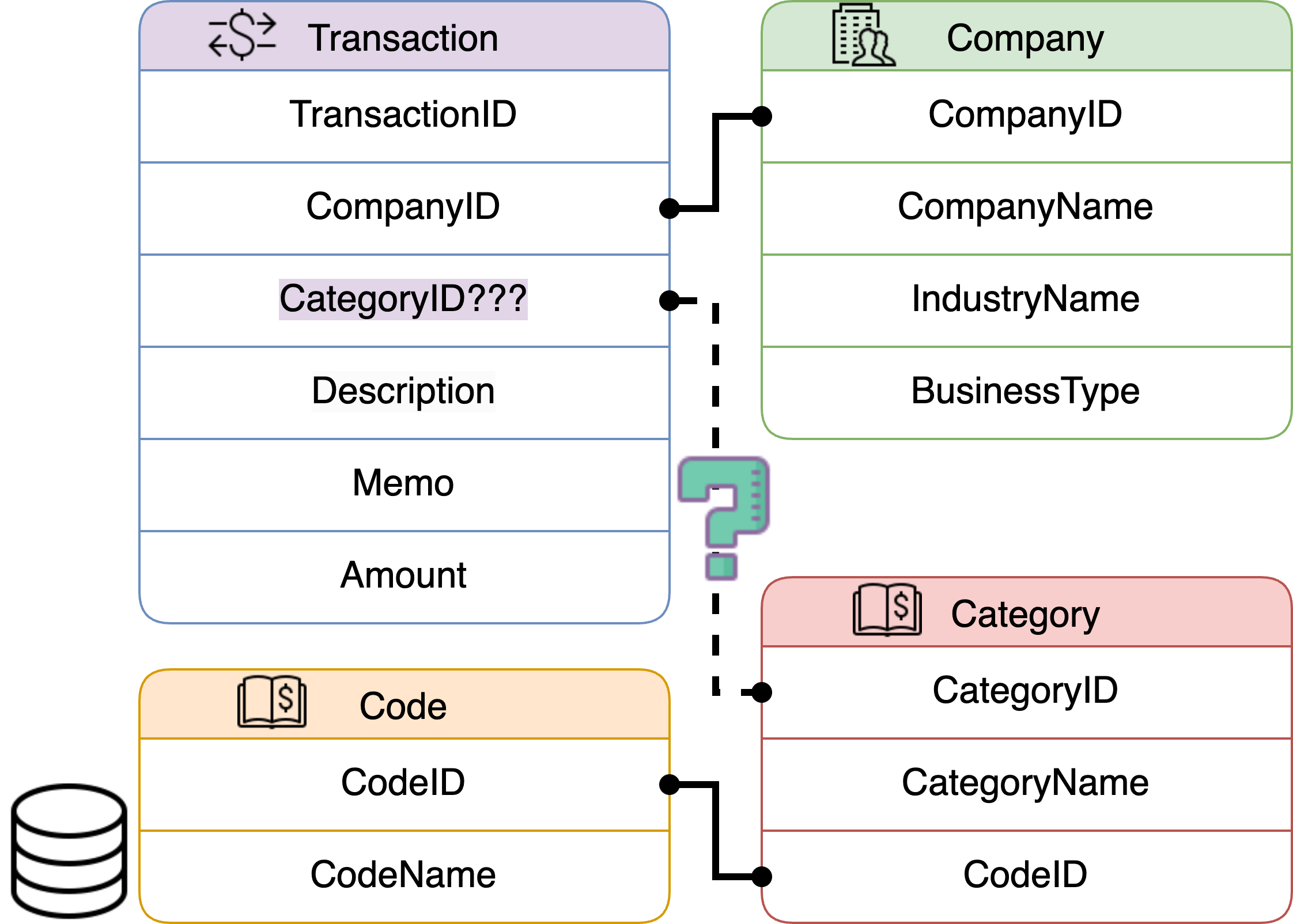}
    \subcaption{\label{fig:schema_test}}
    
\end{subfigure}
\begin{subfigure}[t]{0.28 \linewidth}
    \centering
    \includegraphics[width=\linewidth]{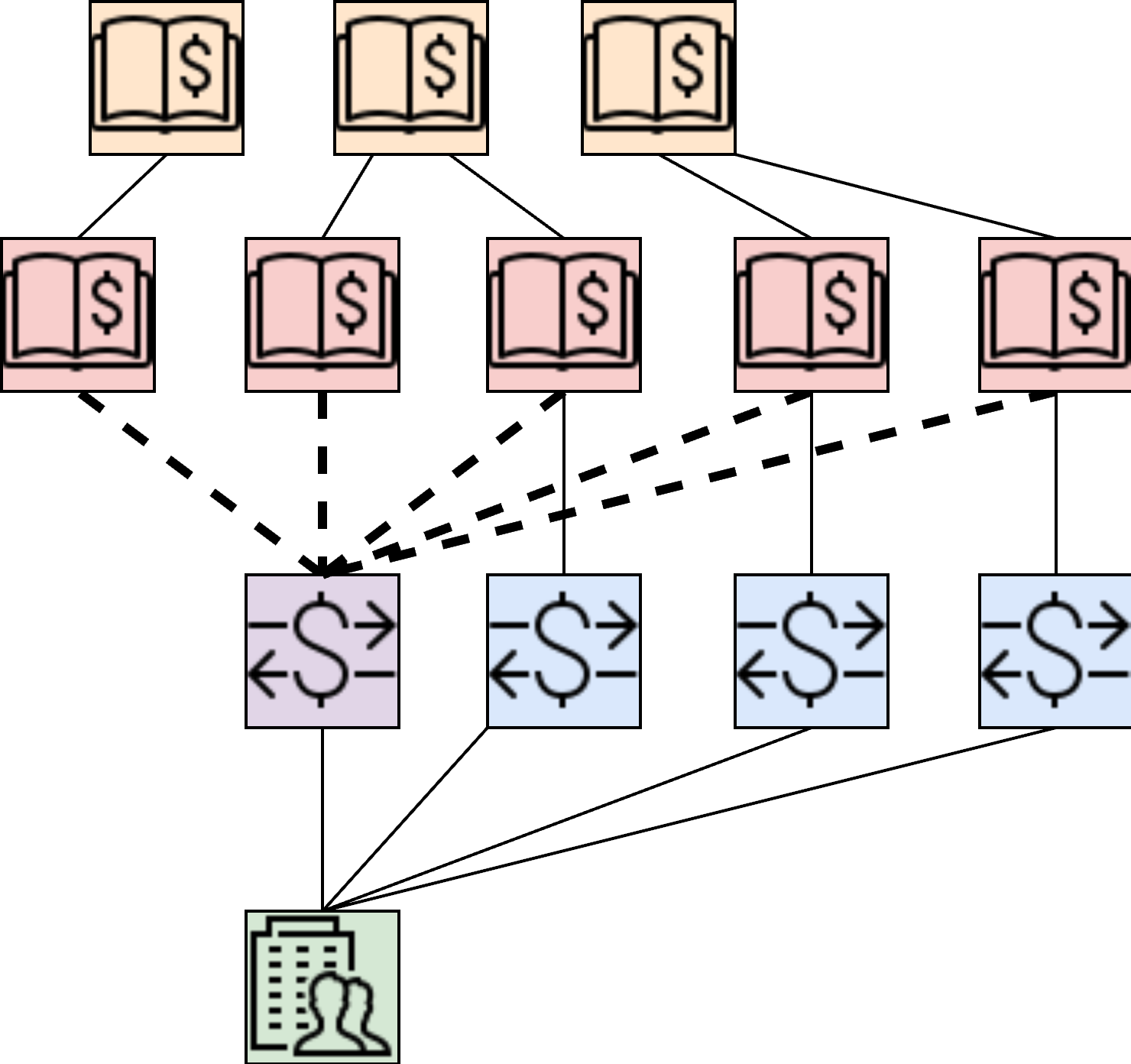}
    \subcaption{\label{fig:to_graph_simple}}
\end{subfigure}
\hspace{2mm}
\begin{subfigure}[t]{0.28 \linewidth}
    \centering
    \includegraphics[width=\linewidth]{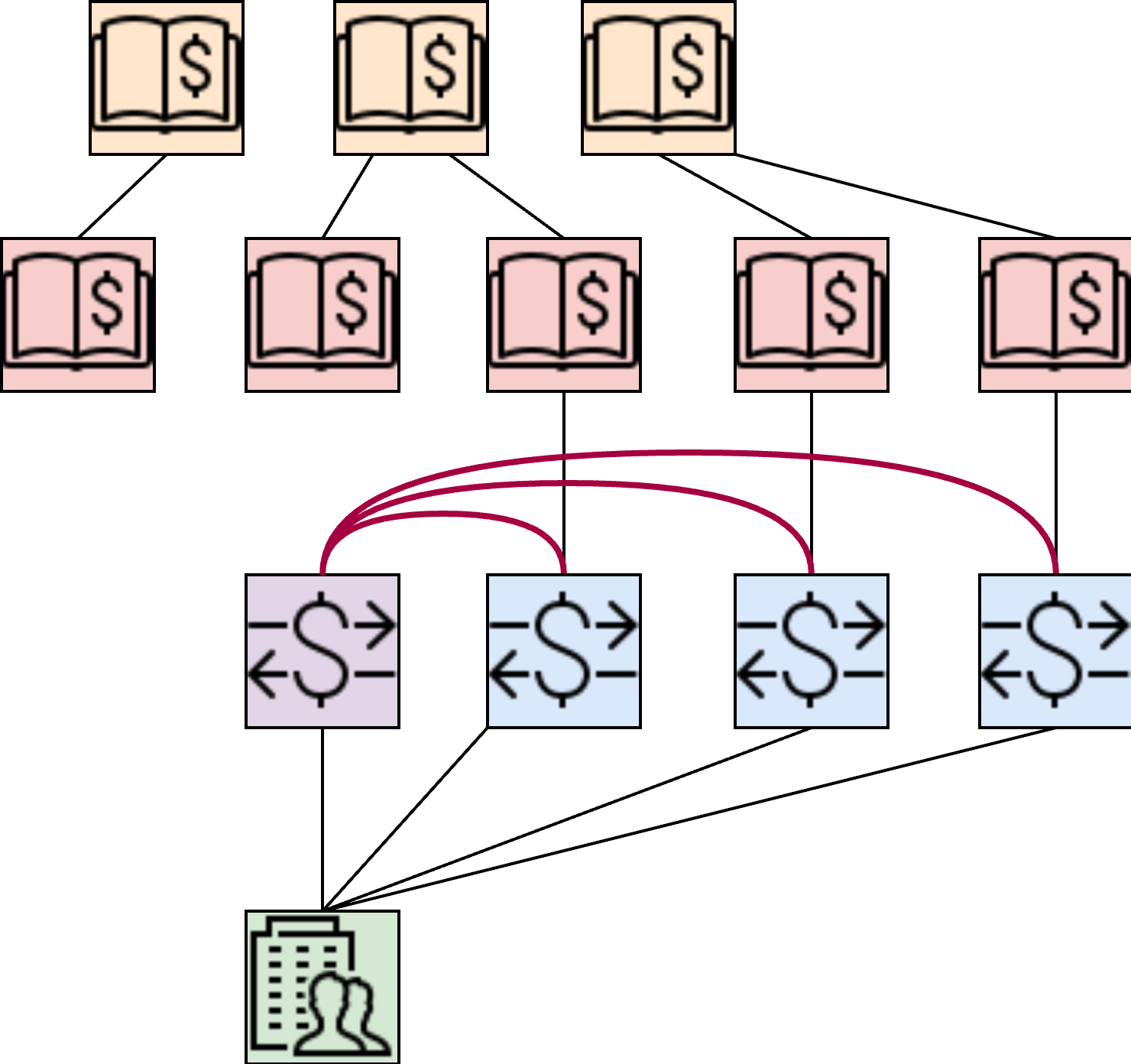}
    \subcaption{\label{fig:to_graph}}
    
\end{subfigure}
\caption{Transformation of a relational database into a heterogeneous graph for transaction categorization. (a) A new transaction enters the system without a foreign key connection to the \cat table. (b) The heterogeneous graph is built from the relational database, where transaction categorization is formulated as a link prediction task. (c) Two-hop connections for transaction nodes mitigate over-squashing and improve model expressiveness.}\label{fig:convert}
\vspace{-6mm}
\end{figure*}

In the heterogeneous graph we've constructed, historical transaction nodes $v \in T_{\text{transaction}}$, that have already been categorized form links with corresponding  \cat nodes, such that $p_{v'} \in \mathcal K_v$ where $v' \in T_{\text{\cat}}$. However, new transactions that have not yet been categorized enter the graph without any existing links to \cat nodes ($p_{v'} \notin \mathcal K_v$ where $v' \in T_{\text{\cat}}$).
This scenario redefines the transaction categorization task. Instead of assigning a \cat to each new transaction, our task shifts to predicting which \cat node in the graph should be linked to the new transaction node. Essentially, the categorization challenge is transformed into a link prediction task as shown in Figure~\ref{fig:schema_test}, where the objective is to determine the most appropriate connections for uncategorized transaction nodes based on the graph's existing structure and the attributes of its nodes. This design also allows us to address the cold start problem, a key challenge for our task, by mapping a transaction from a new user to a relevant \cat based on the historical similarity from other similar businesses.

\subsection{Model the heterogeneous graph}

This section details how \our models the heterogeneous graph $G$. We employ a message-passing Graph Neural Network (GNN)~\cite{gilmer_neural_2017}, specifically a variant of GraphSAGE~\cite{hamilton_inductive_2018}, to encode node representations.
Given the graph's diverse node and edge types, we adapt our approach to model message-passing along different edge types separately. Each message type is initially processed through a homogeneous GNN.
Messages from various edge types are then combined using a second-level aggregation function to get a comprehensive node representation:
\begin{align}
\label{eq:hmp}
   \mathbf{h}^{(i+1)}_v =& t_{\phi(v)}\Bigl(\mathbf{h}^{(i)}_v, \text{AGG}_\text{\small Heter}(\{ \text{AGG}_\text{\small Homo}(\{g_{R}(\mathbf{h}^{(i)}_w) \mid \\ \nonumber
   &w \in \mathcal{N}_R(v)\}) \, \Big\vert \, \forall R=(T,\phi(v)) \in \mathcal{R}\})\Bigr)\textnormal{,} \nonumber
\end{align}
where $\mathcal{N}_R(v) =\{w\in \mathcal{V} \mid (w,v)\in \mathcal{E} \textnormal{ and } \psi(w, v) = R \}$ is the neighborhood of node $v$ under the specific edge type $R$. Here, $\text{AGG}_{\text{Homo}}$ employs a mean operator to normalize message contributions within the same type, while $\text{AGG}_{\text{Heter}}$ uses attention mechanism~\cite{vaswani_attention_2017} to dynamically weight the importance of different message types, improving the model's prioritization of relevant messages

\subsubsection{Two-hop connections for transaction nodes}\label{sec:two_hop}
Our initial node representation learning framework in \our, as outlined in~\autoref{eq:hmp}, aggregates messages from immediate neighbors within the graph $G$. While effective, this approach can introduce issues inherent in GNNs such as over-squashing~\cite{alon_bottleneck_2021} and limited expressiveness~\cite{xu_how_2018}, which we address through graph data augmentation.

\paragraph{Graph data augmentation.}
When GNNs learn the node representation of the \textcolor{txn_test}{target transaction} in~\autoref{fig:to_graph_simple}, they have to propagate at least two rounds of message passing to receive the information from the \textcolor{txn_historical}{historical transactions} of the same company. To enhance node representation, especially for the target transaction node, we introduce an additional type of edge within transaction nodes, effectively making historical transactions direct neighbors of the target transaction. This adjustment allows the target transaction to aggregate messages from historical transactions in just one message-passing iteration. The augmented graph with these new connections is shown in~\autoref{fig:to_graph}. For this new edge type (transaction to transaction), we utilize the GATv2~\cite{brody_how_2022} as the aggregation operator ($\text{AGG}_\text{\small Homo}$). While applying GATv2 to all edge types can lead to GPU memory issues, using it for this specific edge type does not significantly increase memory requirements.

Specifically, we introduce an extra set of edges $\mathcal{E}_{\text{aug}}$ into the original graph: 
\begin{align*}
    \mathcal{E}_{\text{aug}} = \{(v_1, v_2) \in \mathcal{V} \times \mathcal{V} \mid  v_1,v_2 \in T_{\text{transaction}} \text{, } \exists v_c \in T_{\text{company}} \text{, } p_{c} \in \mathcal{K}_{v_1} \cap \mathcal{K}_{v_2}\}.
\end{align*}
The edge set $\mathcal{E}_{\text{aug}}$ is the set of transaction pairs if they are from the same company. Next, we discuss how this modification can mitigate the two issues.

\paragraph{Addressing over-squashing.} GNNs learn the node representation iteratively from the local neighborhood. However, due to this recurrent learning paradigm, GNNs often face a challenge where the information from a growing receptive field is compressed into a fixed-length vector, potentially losing valuable information. This phenomenon is prevalent in the transaction graph $G$. For the \textcolor{txn_test}{target transaction} in~\autoref{fig:to_graph_simple}, the message from its \textcolor{txn_historical}{historical transactions} is equally compressed into one fixed-length vector. Since this message is not conditioned on the target transaction, it cannot adjust itself so that the signals more important to the target transaction are preserved. By rewiring the graph and applying a weighted message aggregator like GATv2, it can effectively enable the target transaction's incoming message to keep the most relevant information~\cite{alon_bottleneck_2021}, enhancing the effectiveness of the prediction.

\paragraph{Enhancing expressiveness.} GNNs essentially simulate the Weisfeiler-Lehman graph isomorphism algorithm~\cite{xu_how_2018}. This limits their expressiveness in terms of distinguishing non-isomorphic graphs. By directly connecting two-hop neighbors (historical transactions) as immediate neighbors to the target transaction in graph $G$, our model approximates a K-hop GNN approach~\cite{feng_how_2023}. This augmentation not only improves expressiveness but does so without significantly increasing computational costs, thus maintains a balance between accuracy and efficiency.

\subsection{Training objective}
The transformation of the relational database into the heterogeneous graph $G$ redefines our task as a link prediction challenge. In essence, link prediction in this context is a ranking problem where the model is expected to rank the correct node pair (the ground truth link) higher than other node pairs (non-connected links). Specifically, for a transaction node $v_i$ and its corresponding \cat node $v_j$ in graph $G$, the score of $(v_i,v_j)$ should be higher than that of any other \cat node pair $(v_i,v_k)$, where $v_k$ represents any \cat node not connected to $v_i$.
We utilize the inner product as the scoring function between transaction and \cat nodes, employing AUCLoss~\cite{wang_pairwise_2022}, a surrogate for AUC, as the training objective of \our:
\begin{equation}
\label{eq:loss}
    \mathcal{L} = \sum_{{(v_i, v_j)\in \mathcal{E}^+, (v_i, v_k)\in \mathcal{E}^-}}\left(1 -  \left(\mathbf{h}_i*\mathbf{h}_j\right) + \left(\mathbf{h}_i*\mathbf{h}_k\right)  \right)^2,
\end{equation}
where $\mathbf{h}_v$ denotes the final representation of either a transaction or \cat node from~\autoref{eq:hmp}. The set of positive node pairs $\mathcal{E}^+$ is chosen as:
\begin{equation}\label{eq:pos}
\mathcal{E}^+ \subseteq \{(v_i, v_j) \in T_{\text{transaction}} \times T_{\text{\cat}} \mid p_{v_j} \in \mathcal{K}_{v_i}\}.
\end{equation}
The set of negative node pairs, $\mathcal{E}^-$, includes non-connected links, defined as:
\begin{equation}
\mathcal{E}^- \subseteq \{(v_i, v_k) \in T_{\text{transaction}} \times T_{\text{\cat}} \mid p_{v_j} \notin \mathcal{K}_{v_i}\}.
\end{equation}

\subsubsection{Weighted negative sampling}\label{sec:weighted_neg_sampling}
The set of negative node pairs, $\mathcal{E}^-$, consists of non-connected links. 
Due to the impracticality of enumerating all potential negative pairs, we employ a negative sampling strategy. For a transaction node $v_i$ and its corresponding positive \cat node $v_j$, simple uniform sampling from $T_{\text{\cat}} \setminus \{v_j\}$ would be suboptimal due to the long-tail distribution of \cat. This could prevent effective learning if those \cat nodes $v_k$, which rarely appear as positives, are sampled as negatives. To address this, we employ a weighted multinomial distribution for negative sampling, where weights are assigned proportional to the normalized frequency of \cat appearances, enhancing the relevance and challenge of the sampled negatives.

\subsubsection{Selective Loss Computation via Diversity Filtering}\label{sec:diverse-filtering}
During the loss computation and backpropagation over a batch - not all samples contribute equally to learning, some being too easy (providing little gradient signal) or redundant (leading to inefficient updates). Thus, we propose a selective loss computation strategy, where all samples undergo a forward pass, but only a subset of challenging or informative samples contribute to the loss and gradient computation, i.e., a full-forward partial-backward strategy. This strategy retains the benefits of large-batch inference while ensuring that gradient updates focus on samples that are most challenging and diverse. Since all samples contribute to feature extraction during the forward pass, the model benefits from seeing a broader distribution of the data. 
This results more accurate hard sample identification, targeted gradient updates, and improved generalization. 
To select the subset of diverse samples, we compute the dot product similarity of transaction feature representation. From this distribution, we gradually refine the selection over epochs, choosing samples with lower similarity scores and only use those for loss computation. Initially, we utilize 100$\%$ of the samples, then gradually decrease this proportion in stages until reaching 40$\%$ , maintaining a core set of diverse samples throughout training. Experiments show that this provides us with the optimal diversity samples for loss computation in a batch, providing highest gradient signals during back-propagation. The impact of varying extents of diversity sampling is analyzed in Appendix\ref{app:supplementary}. 

\subsection{Scalability and practical designs}
In this section, we detail several scalability improvements to \our, designed to effectively manage the extensive volume of transaction data encountered in real-world applications. We discuss practical designs to reduce neighborhood sizes for transaction and \cat nodes during node representation encoding and introduce a rule-based early exit method, Top-K Nearest Neighbor, to efficiently manage the workload on GNNs.

\subsubsection{Reducing neighborhood size}
Given that GNNs encode node representations by aggregating features from local neighborhoods, large neighborhoods can pose significant scalability challenges, particularly in terms of memory consumption. To address this, we implement a node-wise neighbor sampling strategy. Different from existing methods~\cite{hamilton_inductive_2018,frasca_sign_2020,ying_graph_2018}, we further take node types into consideration to optimize the receptive field of the GNNs. We employ distinct strategies for transaction nodes and \cat nodes:

\paragraph{Similarity-based neighbor sampling for transaction nodes.}\label{sec:similarity_sampling}
For transaction nodes, $\{v \mid v \in T_{\text{transaction}}\}$, the local neighborhood typically includes all historical transactions associated with the company, which can be extensive. For example, there are companies owning over $2000$ transactions within just one year. To manage this, we utilize similarity-based neighbor sampling, which reduces neighborhood size while preserving relevant information.

We compute the cosine similarity between the text embeddings of the target transaction node and its historical counterparts. Historical transactions are then sampled based on their similarity scores, prioritizing those most relevant to the target. This approach not only constrains computational overhead but also ensures that the most informative connections are maintained in \our's neighborhood. Techniques like Faiss~\cite{douze_faiss_2024} can be employed to enhance the efficiency of these computations.

\paragraph{Edge direction dropping for \cat nodes.}
\cat nodes, $\{v \mid v \in T_{\text{\cat}}\}$, often have neighborhoods that include a huge set of transaction nodes linked to them. The popularity of some \cat can lead to extremely large neighborhood sizes, which are not only impractical to process, but also ineffective to model.

To mitigate this, we discard all incoming edge connections from transaction to \cat nodes. This adjustment significantly reduces the computational load by limiting the receptive field to exclude transaction nodes, while outgoing connections from \cat to transaction nodes are kept. This approach not only simplifies the computation required to encode \cat node but also ensures consistency in \cat node representation during inference, as the computational graph remains unchanged regardless of new transactions being added.

\subsubsection{Top-K Nearest Neighbor}
Despite the sophisticated capabilities of \our, equipped with GNNs to perform predictions based on graph structure $G$, real-world scenarios often present varying degrees of prediction difficulty~\cite{panda_conditional_2016}. Many transactions imported into \QB by users are very similar or even identical to past entries due to routine business activities. In such cases, users frequently reuse the same \cat as in previous transactions. Consequently, a significant portion of transactions could be accurately categorized without necessitating full GNN processing. To capitalize on this, we utilize the streamlined and effective early exit method through Top-K Nearest Neighbor (\topknn).

Upon the arrival of a new transaction, while we still prepare the graph $G$ for subsequent GNN processing, we first assess if sufficiently similar historical transactions might already provide reliable predictions. Utilizing the similarity scores computed during the similarity-based neighbor sampling of transaction nodes, we identify a subset of historical transactions that exhibit a similarity score exceeding a cutoff, set at $0.8$ for our experiments.

From this subset, we directly derive \cat from the top-$K$ most similar transactions. Given that our categorization task aims to predict the $5$ most likely categories, \our will output these labels directly if $5$ distinct \cat are available from this subset. If fewer than $5$ categories are available, the graph $G$ is then processed by the GNNs to generate the remaining predictions.
\section{Experiments}

\subsection{Experimental setup}

\subsubsection{Dataset}
To evaluate the performance of \our comprehensively, we curated a dataset from active \QB users as of November 2023. We randomly selected $7.5K$ companies and used their two most recent transactions with labeled \cat post-November 2023 as our test set, resulting in a total of $15K$ transactions. The training set consisted of $3000K$ transactions having $100K$ \cat across $15K$ companies reflecting a wide range of user preferences and patterns. 

\subsubsection{Experimental settings}
\label{sec:setup}
We benchmark the performance of \our against the current production models in \QB, namely \texttt{Shorthair} and \texttt{Lynx}. Shorthair is a population model that employs contrastive learning and Word2Vec embeddings, whereas Lynx, built on top of Shorthair is a logistic regression model customized to a company. We assess the models using the metrics of Top-1, Top-2, and Top-5 accuracy, which measure whether the correct label is among the Top-k predictions of the model, ranked by the \cat scores. We conduct evaluations under two distinct settings:

\textit{Zero Shot: }In this setting, categorization is based solely on the information from the new transaction itself, without any contextual data from the owning company or its historical transactions. Both \texttt{Shorthair} and \bert are evaluated under this setting.

\textit{Few Shot: }This setting incorporates not only the data from the new transaction but also contextual information from the owning company and its historical data. \texttt{Lynx}, \topknn, and \our are assessed under this framework.

\subsection{Results}
\sloppy
The experimental results are presented in~\autoref{table:main}. In the Zero Shot setting, \bert outperforms the production \texttt{Shorthair} model significantly. The \bert model with 6 layers achieves a Top-1 accuracy boost of $7.76\%$ and a Top-5 accuracy of $74.12\%$, indicating that our trained-from-scratch text encoder effectively captures the semantics of transaction descriptions and maps them accurately to the corresponding \cat. The 12-layer \bert model shows only marginal improvement over the 6-layer model, suggesting that a lightweight language model pretrained from scratch is sufficient for encoding transaction data.

\begin{wraptable}{r}{0.5\textwidth}
\caption{Transaction categorization evaluated by accuracy under Zero Shot and Few Shot settings.}
\label{table:main}
  \resizebox{1\linewidth}{!}{
  \begin{NiceTabular}{lccc}
    \toprule
    \textbf{Methods} & \textbf{Top-1} & \textbf{Top-2} & \textbf{Top-5} \\
    
    \midrule
    \rowcolor{Gray}
    \multicolumn{4}{c}{\emph{Zero Shot}} \\
    \texttt{Shorthair} & 36.07 & - & - \\
    \bert (6 layers) &  43.83 & 57.96 & 74.12 \\
    \bert (12 layers) &  \textbf{45.52} & \textbf{59.47} & \textbf{75.46}\\
    
    \midrule
    \rowcolor{Gray}
    \multicolumn{4}{c}{\emph{Few Shot}} \\
    \texttt{Lynx} & 62.49 & - & - \\ 
    \topknn & 65.80 & 73.63 & 78.55 \\
    \our (GNNs only) & 63.38 & 74.60 & 84.89 \\ 
    \our &  \textbf{68.67} & \textbf{78.97} & \textbf{88.04} \\ 

    \midrule
    \rowcolor{Gray}
    \multicolumn{4}{c}{\emph{Ablation Study}} \\
    \our (GNNs only) & 63.38 & 74.60 & 84.89 \\
    w/o \bert & 55.46 & 64.02 & 73.16 \\
    w/o two-hop connections & 47.07 & 61.16 & 76.58 \\
    w/o similarity sampling & 56.39 & 67.89 & 80.63 \\
    w/o diversity filtering & 61.74 & 73.18 & 83.85 \\
  \bottomrule
\end{NiceTabular}
\vspace{-5mm}
}
\vspace{-5mm}
\end{wraptable}

In the Few Shot setting, \our demonstrates substantial performance advantages. It achieves a Top-1 accuracy of $68.67\%$ and a Top-5 accuracy of $88.04\%$, significantly outperforming the \texttt{Lynx} model. Additionally, the~\topknn method, which uses text embeddings from \bert, achieves a Top-1 accuracy of $65.80\%$, surpassing the performance of \our with GNNs alone. This result highlights the effectiveness of \topknn in identifying recurring transactions in a user's history, thereby enhancing Top-1 accuracy of~\our.

However, when the transaction categorization must be inferred beyond the user's most similar historical transactions, the GNN component of \our exhibits superior generalizability, achieving nearly $85\%$ in Top-5 accuracy. This demonstrates that while \topknn is highly effective for repeated transactions, the GNN module of \our provides a broader and more accurate categorization capability for diverse transaction scenarios.



\subsection{Seen vs Unseen Category in a Company's History}

In this section, we discuss why \our is designed as a hybrid model, namely a combination of \topknn and GNNs. In Table~\ref{table:hs_hu}, we not only report our overall accuracy, but further break down the performance of \our into Historical Seen and Historical Unseen scenarios. For Historical Seen, we choose the test samples such that their ground-truth \cat is present within the company's own history as context. For Historical Unseen, we choose the samples whose \cat are present in the overall dataset but unseen to that company's history. 

We note that \topknn, while having the best performance at repeated labels, has no predictive power for unseen labels. For this subset, \our's GNN is able to detect the labels with over 22$\%$ accuracy, highlighting its generalizable ability. Given that most transactions in production systems belong to the historical seen category, \our has the best overall performance, balancing between both the subsets, thereby demonstrating the need for GNN and \topknn hybrid model. Furthermore, analyzing these trends over the ablation studies, we note that each design choice adds to the overall accuracy, augmenting performance for either or both the seen and unseen subsets.

\begin{wraptable}{r}{0.5\textwidth}
\caption{Performance breakdown in different scenarios. \textbf{Acc} is the overall Top-1 accuracy, \textbf{HS} is the accuracy on Historical Seen subset, and \textbf{HU} is the accuracy on Historical Unseen subset.}
\label{table:hs_hu}
\centering
  \resizebox{1\linewidth}{!}{
  \begin{NiceTabular}{l ccc}
        \toprule
        \textbf{Methods} & \textbf{Acc} & \textbf{HS} & \textbf{HU} \\
        \midrule
        \topknn &  65.80 & \textbf{79.72} & 0.16 \\
        \our (GNNs only) & 63.38 & 72.25 & \textbf{22.05} \\ 
        \our &  \textbf{68.67} & \underline{78.64} & \underline{20.84} \\
        \midrule
        \textbf{Methods (GNNs only)} & \textbf{Acc} & \textbf{HS} & \textbf{HU} \\
        \midrule
        \our (GNNs only) & 63.38 & 72.25 & 22.05 \\
        w/o \bert &  55.46 & 66.24 & 4.83\\
        w/o two-hop connections & 47.07 & 51.36 & 26.93\\
        w/o similarity sampling & 56.39 & 65.26 & 14.73 \\
        w/o diversity filtering & 61.74 & 70.46 & 20.75 \\
        \bottomrule
    \end{NiceTabular}
    }
    \vspace{-5mm}
\end{wraptable}

\vspace{12mm}
\subsection{Ablation Studies}

In this section, we delve deeper into the validation of the effectiveness of various design choices in \our. Additional supplementary experiments are in Appendix~\ref{app:supplementary}.

In~\autoref{table:main}, we report ablation studies to validate if the proposed components in \our can enhance the predictive power of transaction categorization. For the ablation studies, we only include the GNNs module for \our.

\sloppy
\textit{w/o \bert (Sec~\ref{sec:txn-bert}): }Replacing \bert with off-the-shelf Sentence-Bert~\cite{reimers_sentence-bert_2019}, we observe a decrease in performance, underscoring the necessity of a trained-from-scratch text encoder tailored to transaction data. We further note the steep drop in performance for the unseen subset.

\textit{w/o two-hop connections (Sec~\ref{sec:two_hop}): }This ablation study emphasizes the critical role of explicit graph augmentation. It demonstrates that directly connecting the target and historical transactions as neighbors significantly enhances performance. This finding underscores that merely converting a relational database to a heterogeneous graph without strategic modifications is inadequate for maximizing the model's effectiveness. While this setting has the best performance for the unseen subset, it comes at the expense of significant decline in overall performance.

\textit{w/o similarity sampling (Sec~\ref{sec:similarity_sampling}): }Testing the impact of neighbor sampling based on semantic similarity, we find that sampling similar historical transactions in GNN's computation graphs allows \our to leverage relevant information effectively for accurate predictions. We note that this setting also suffers from lack of generalizability.

\textit{w/o diversity filtering (Sec~\ref{sec:diverse-filtering}): }here we report the results without the diversity filtering, when all samples pass through the backward pass during GNN training throughout the training. We note that while the drop in the unseen subset performance is to be expected without this filtering, we also note a drop across all metrics, suggesting the effectiveness of selective backpropagation. 

\subsection{Time complexity}

We have assessed the processing time for various components within \our, with the results detailed in~\autoref{table:time}. We report the processing time as time taken in milliseconds (ms) per $1,000$ transactions. The entire pipeline of \our is designed to operate effectively on both CPU and GPU environments, catering to different production system requirements.
\begin{wraptable}{r}{0.35\textwidth}
\vspace{-6mm}
\caption{Inference times for $1,000$ transactions.}
\label{table:time}
\centering
  \resizebox{1\linewidth}{!}{
  \begin{NiceTabular}{l|cc}
    \toprule
    \textbf{Walltime (ms)} & \textbf{CPU} & \textbf{GPU}\\

    \midrule
    \bert & 1400 & 67\\
    \our & 6808 & 843\\
    - \topknn & 333 & - \\ 
    - \our (GNNs only) &  6614 & 324\\
    
    \midrule
    \emph{Total} & 8208 & 910\\
  \bottomrule
\end{NiceTabular}
}
\vspace{-8mm}
\end{wraptable}
On a CPU, the lightweight text encoder \bert processes $1,000$ transactions in just $1400$ milliseconds. To deliver the Top-5 predictions, \our requires approximately $8$ seconds for $1,000$ transactions. In contrast, utilizing a GPU significantly reduces the processing time to under $1$ second for outputting the Top-5 predictions. This efficiency demonstrates \our's capability to scale and meet the demands of real-world transaction volumes effectively.

\subsection{Prediction cascade}
\begin{figure}[h]
    \centering
    \resizebox{0.5\linewidth}{!}{%
    \includegraphics[width=1.0\linewidth]{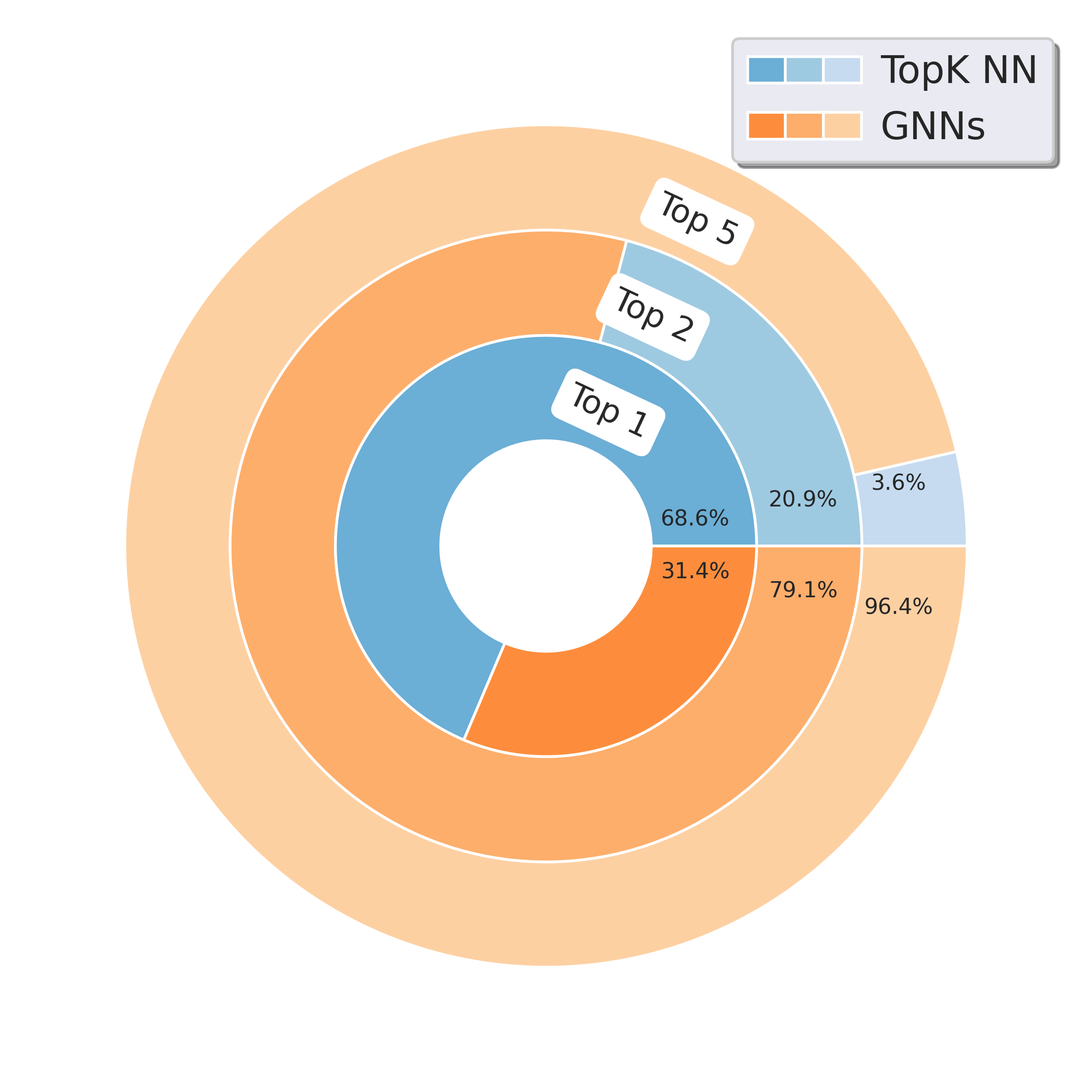}
    }
    \caption{Cascade process in \our. TopK NN efficiently resolves $68\%$ of transactions when only a Top-1 prediction is needed. However, for more comprehensive Top-5 predictions, over $96\%$ of transactions necessitate processing by GNNs.}
    \label{fig:cascade}
\end{figure}

In the \our pipeline shown in~\autoref{fig:framework}, the processed graph initially enters the TopK NN module, which serves as an early exit strategy to identify similar transactions via text embeddings. If this step does not yield sufficient \cat predictions, the pipeline advances to the GNNs for additional predictions.

Our analysis of this cascade, shown in~\autoref{fig:cascade}, reveals that for Top-1 predictions, TopK NN alone efficiently handles over $68\%$ of transactions. This high percentage underscores the prevalence of similar transactions within the database. However, to generate the Top-5 predictions, more than $96\%$ of transactions require processing by the GNNs, indicating the need for a more in-depth computation to fulfill broader prediction requirements.

This flexible approach in \our demonstrates the system's adaptability, allowing for a balance between efficiency and thoroughness in prediction based on system demands and user experience considerations.
\section{Conclusion}
In this study, we introduce \our, a unified model designed for transaction categorization within \QB. Recognizing the unique linguistic characteristics of transaction data, we train, from scratch, a \bert text encoder, to grasp the semantic nuances of financial transaction descriptions. Subsequently, we employ a GNN-based model to capture the relationships among the tables in a relational database, redefining transaction categorization as a link prediction task over a heterogeneous graph. To address the challenges posed by the high volume of transactions and the specific demands of the categorization task, we integrate several innovative components that significantly boost \our's predictive capabilities. Our experimental results demonstrate that \our not only surpasses previous production models in terms of performance but also offers remarkable scalability and efficiency, making it well-suited for handling large-scale transaction data in real-world settings. Having proved its merit in offline testing, our approach is currently in the process of being implemented for deployment in production due to its improved metrics, improved customer experience providing multiple options for categories to choose from (top-$k$), and the simplified overall architecture allowing the company to move away from running and maintaining million plus models in production for supporting personalized transaction categorization.

\begin{credits}
\subsubsection{\ackname} We thank the anonymous reviewers for their insightful comments and helpful discussions. We are also grateful to Byron Tang, Heather Simpson, Jocelyn Lu, and Hilaf Hasson for their assistance in retrieving the dataset and providing constructive feedback on this project.
\end{credits}

\bibliographystyle{splncs04}
\bibliography{output2}

\newpage
\appendix
\section{Technical details}
\subsection{Custom tokenizer}\label{app:tokenizer}
Transaction descriptions, by their nature, resemble a constructed language distinct from natural language due to the structured format. This unique format is filled with variations of business entity names and their abbreviations, necessitating a customized vocabulary for effective text processing.

To create this specialized tokenizer, we start by gathering all available transaction descriptions to form a dedicated corpus. We employ the WordPiece tokenization method~\cite{schuster_japanese_2012}—similar to the one used in the BERT model~\cite{wu_googles_2016}—but develop a unique vocabulary tailored to the transaction descriptions.


\begin{figure*}[h]
    \centering
\begin{subfigure}[t]{0.65 \textwidth}
    \includegraphics[width=\linewidth]{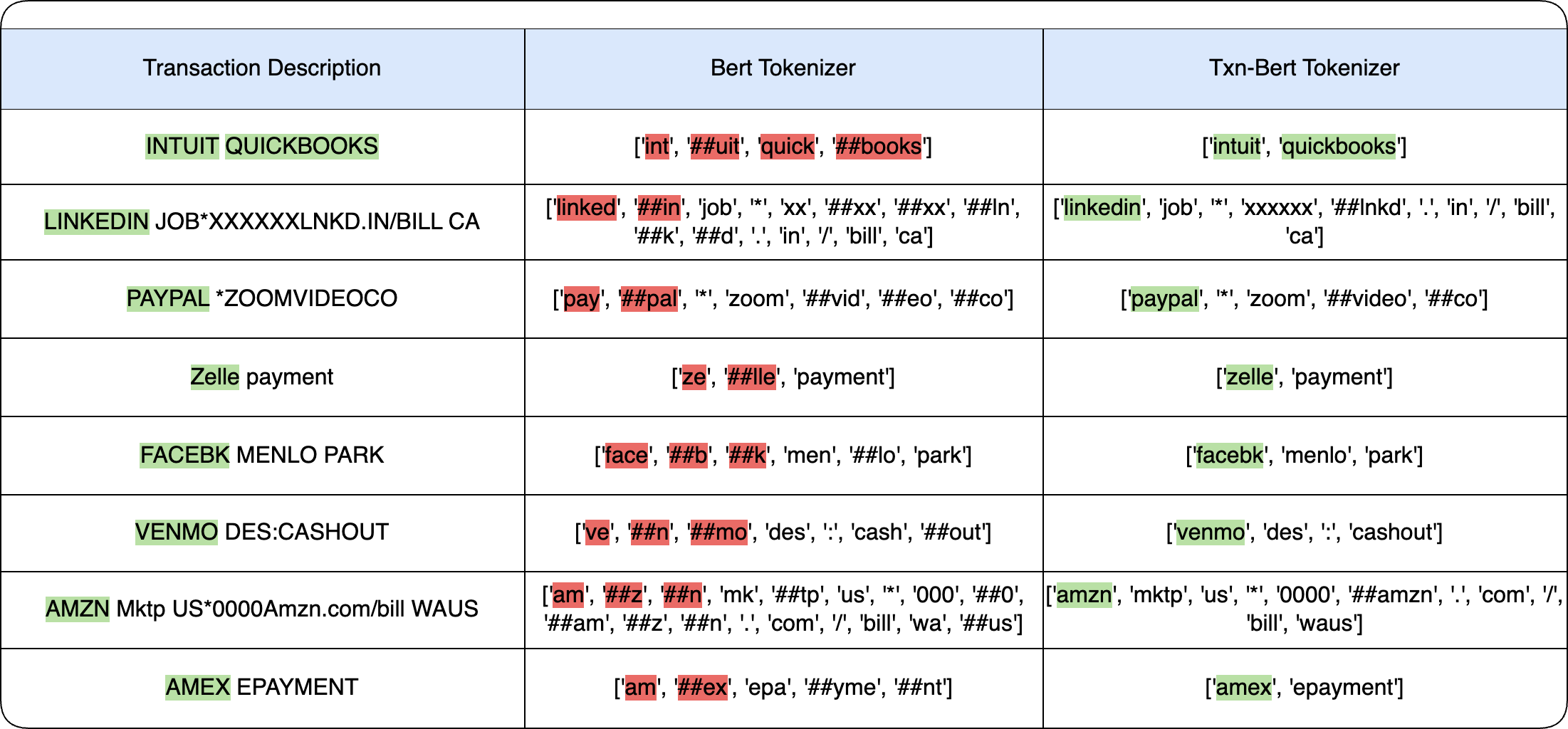}
    \subcaption{\label{fig:tokens}}
    
\end{subfigure}%
\begin{subfigure}[t]{0.35 \textwidth}
    \centering
    \includegraphics[width=\linewidth]{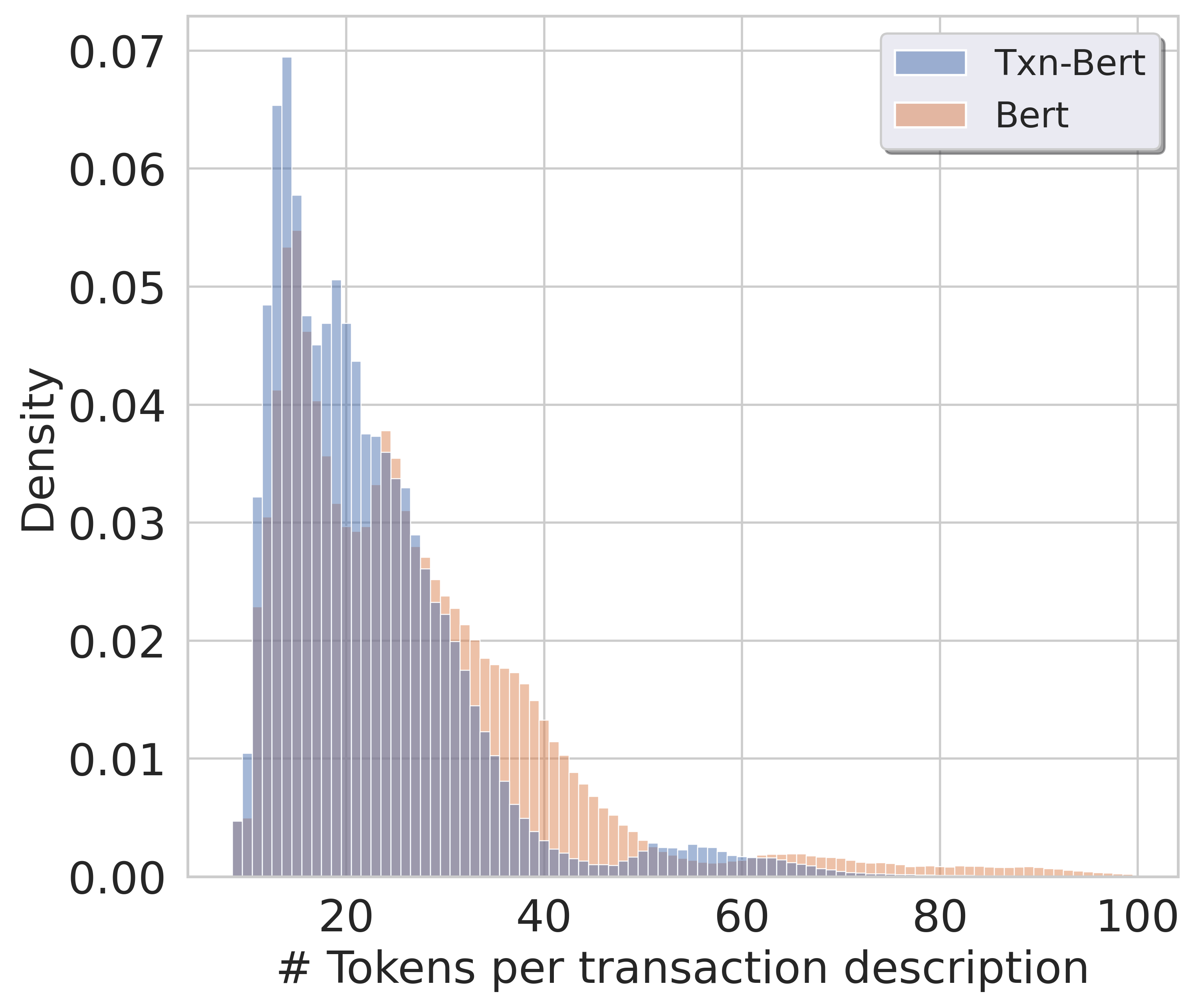}
    \subcaption{\label{fig:tokens_dist}}
    
\end{subfigure}
\caption{Tokenization of transactions by BERT and \bert. (a) Bert tokenizer tends to split business entity names to sub-tokens, while \bert tokenizer can keep them complete. (b) \bert can tokenize transactions into fewer tokens.}\label{fig:tokenizer}
\end{figure*}

~\autoref{fig:tokenizer} compares the tokenization of transaction descriptions between the standard BERT model and our pretrained \bert. As illustrated in ~\autoref{fig:tokens}, the standard BERT tokenizer often breaks down business names into smaller pieces, whether abbreviated or not. This occurs because these business names, despite being prevalent in transaction descriptions, are not included in the standard BERT vocabulary, which is designed for general natural language. In contrast, our custom-built tokenizer, which is trained specifically on transaction data, recognizes complete business names as single tokens. This capability not only simplifies the text representation but also allows for more efficient data processing, as shown in~\autoref{fig:tokens_dist}. This indicates that our trained-from-scratch tokenizer can capture the entire transaction descriptions with fewer bits and represent the text distribution better. Further empirical experiments in Appendix~\ref{app:ablation} show that a custom tokenizer can improve language modeling performance on transaction data.

\subsection{\bert configuration}\label{app:bert_config}
For the text encoding within \our, we utilize the transformer architecture from Sentence-Bert~\cite{reimers_sentence-bert_2019}, specifically leveraging the HuggingFace implementation of Sentence-Bert\footnote{https://huggingface.co/sentence-transformers/all-MiniLM-L6-v2}. We primarily employ a 6-layer transformer configuration for \our due to its efficiency and sufficiency in capturing relevant features from the transaction data. In our experiments, as detailed in~\autoref{table:main}, a 12-layer transformer was tested solely for ablation study purposes. The results indicated that the 6-layer model provided comparable performance with marginal benefits from the more complex 12-layer model.



\subsection{Details of converting databases to graphs}
\label{app:convert}
Adopting the notations from~\cite{fey_position_2024}, we define the relational database as $(\mathcal{T}, \mathcal{L})$, consisting of a collection of tables $\mathcal{T} = \{T_i\}$ where $i$ represents different tables such as ``transaction'', ``company'', ``\code'', or ``\cat''. The relationships between these tables are captured by links $\mathcal{L} \subseteq \mathcal{T} \times \mathcal{T}$. An edge $L = (T_{\text{fkey}}, T_{\text{pkey}})$ exists if a foreign key column in $T_{\text{fkey}}$ points to a primary key column in $T_{\text{pkey}}$.

Each table $T$ comprises a set of rows $T = \{v_1, \dots, v_{n_T}\}$, where each row $v \in T$ includes three components: (1) Primary key $p_v$ uniquely identifies the row $v$ within its table. (2) Foreign keys $\mathcal K_v\subseteq \{p_{v'}:  v' \in T'\text{ and } (T, T') \in \mathcal L \}$ establish connections to rows $v'$ in other tables $T'$. (3) Attributes $x_v$ hold the informational content of the row, such as textual data, which are encoded using \bert to generate embeddings.

\sloppy
We also define the relation type $\mathcal{R} = \mathcal{L} \cup \mathcal{L}^{-1}$, where $\mathcal{L}^{-1} = \{(T_{\text{pkey}}, T_{\text{fkey}}) | (T_{\text{fkey}}, T_{\text{pkey}}) \in \mathcal{L}\}$, representing the inverse of the primary-foreign key links.
The heterogeneous graph is formalized as $G = (\mathcal{V}, \mathcal{E}, \phi, \psi)$:
\begin{enumerate}
    \item $\mathcal{V}$: Node set, representing rows across all tables.
    \item $\mathcal{E} \subseteq \mathcal{V} \times \mathcal{V}$: Edge set, representing relationships based on primary-foreign key mappings.
    \item $\phi: \mathcal{V} \rightarrow \mathcal{T}$: Node type mapping function, assigning each node to its corresponding node type (table).
    \item $\psi: \mathcal{E} \rightarrow \mathcal{R}$: Edge type mapping function, linking each edge to its relation type.
\end{enumerate}

We then map the elements of the relational database $(\mathcal T, \mathcal L)$ to the elements of the heterogeneous graph $G=(\mathcal{V}, \mathcal{E}, \phi, \psi)$. We first define the node set in the converted graph as the union of all rows in all tables $\mathcal{V} = \bigcup_{T\in\mathcal{T}}T$. Its edge set is then defined as
\begin{equation}
\mathcal{E} = \{(v_1, v_2) \in \mathcal{V} \times \mathcal{V} \mid  p_{v_2} \in \mathcal{K}_{v_1} \text{ or } p_{v_1} \in \mathcal{K}_{v_v}\}.
\end{equation}
That is, the edge set is the row pairs that arise from the primary-foreign key relationships in the database. Therefore, the type mapping function can be defined as $\phi(v) = T$ for all $v\in T$ and $\psi(v_1,v_2) = (\phi(v_1), \phi(v_2)) \in \mathcal R$ if $(v_1,v_2) \in \mathcal E$. The attributes $x_v$ hold the node attributes for each node $v$.

\subsection{Inference}
In both the Zero Shot and Few Shot settings, we determine the scores by computing the inner product between the transaction and \cat embeddings to identify the Top-5 most likely predictions for a given transaction.

In Zero Shot setting, \bert encodes both the new transaction and all \cat labels into text embeddings. The cosine similarities between the transaction embedding and each \cat are then calculated, ranked, and the Top-5 predictions are selected based on these rankings.

For the Few Shot setting, Top-K NN is firstly employed to identify similar historical transactions for a given company, with a similarity threshold set at $0.8$. The \cat labels corresponding to these similar historical transactions are directly used as predictions. If fewer than 5 distinct \cat are identified through this method, we engage the GNN component of \our to compute embeddings for the transaction and \cat. A ranking process is then conducted to select any remaining predictions needed to complete the Top-5.

\subsection{Distribution shift mitigation}\label{sec:fakeedge}
\sloppy
During training, the existing edge connections between transaction and \cat nodes in $\mathcal{E}^+$ can potentially create a distribution shift~\cite{dong_fakeedge_2022}. These connections can be inadvertently learned as shortcuts by the GNN, which are not available during testing, thereby affecting model performance.
To mitigate this, we adopt a strategy similar to the FakeEdge method~\cite{dong_fakeedge_2022}, but implement in batch mode. Specifically, in each training epoch, we randomly select approximately $5\%$ of all positive node pairs from $\mathcal{E}^+$ to serve as our positive set $\mathcal{E}^+$. Then, we mask the edges in the graph if the node pairs are from the positive set:

\begin{equation}
    \mathcal{E} := \mathcal{E} \setminus \mathcal{E}^+.
\end{equation}
This separation creates two distinct edge groups within the graph: one that provides the training signals without any direct links ($\mathcal{E}^+$), and another that maintains the graph structure for message passing ($\mathcal{E} \setminus \mathcal{E}^+$). This ensures that during training, the connections used for supervision do not give away the actual links, thus preventing the model from leveraging these as shortcuts and better simulating the conditions it will encounter during testing.



\subsection{Software and Hardware details}
We develop \our using the PyTorch framework, PyTorch Geometric for graph neural network operations, and the HuggingFace library for transformer architectures. All experiments were conducted on an Amazon SageMaker instance equipped with four V100 GPUs.

\section{Supplementary experiments}\label{app:supplementary}

\subsection{More ablation studies}\label{app:ablation}
We conduct more ablation studies to validate the proposed components in the paper. It can be found in~\autoref{table:ablation_appendix}.

\begin{table}[h]
\caption{More ablation study for \our.}
\label{table:ablation_appendix}
\centering
  \resizebox{0.5\linewidth}{!}{
  \begin{NiceTabular}{l|ccc}
    \toprule
    \textbf{Methods} & \textbf{Top-1} & \textbf{Top-2} & \textbf{Top-5} \\
    \midrule
    
    \bert (6 layers) &  43.83 & 57.96 & 74.12 \\
    w/o custom tokenizer & 41.34 & 55.52 & 71.34 \\
    \midrule
    \our (GNNs only) & 63.38 & 74.60 & 84.89 \\ 
    w/o GATv2 on transactions & 50.02 & 63.86 & 79.04 \\ 
    w/o weighted negative sampling & 35.76 & 50.48 & 70.45 \\ 
    w/o distribution shift mitigations & 24.16 & 36.87 & 57.86 \\ 
    w/ stricter diversity sampling & 49.52 & 61.70 & 76.68\\ 
    w/ lenient diversity sampling & 58.27 & 70.81 & 83.21\\ 
    replacing w/ MPNet & 59.06 & 70.22 & 81.75\\ 
  \bottomrule
\end{NiceTabular}
}
\end{table}

\paragraph{w/o custom tokenizer (Appendix~\ref{app:tokenizer})} Standard tokenizer is trained on a corpus significantly different from the transaction data. By training new tokenizer, the tokenization process becomes significantly more efficient (e.g., in terms of average number of tokens), and it avoids unnecessary splitting of important words, making it easier for language modeling. As a result, \bert with custom tokenizer can improve Top 1/2/5 performance by $3$ percent in the Zero Shot setting.


\paragraph{w/o GATv2 on transactions (Sec~\ref{sec:two_hop})}Examining the necessity of GATv2 for transaction-to-transaction message passing, the results indicate that such a GNN structure enhances \our's performance by enabling conditioned message passing.

\paragraph{w/o weighted negative sampling (Sec~\ref{sec:weighted_neg_sampling})}This study evaluates the importance of weighted negative sampling in training \our. Findings suggest that this approach is crucial for achieving optimal performance and addressing the challenges posed by the skewed distribution of \cat labels.

\paragraph{w/o distribution shift mitigations ~\ref{sec:fakeedge}}Assessing our approach to mitigating distribution shift shows that this is a significant issue in link prediction tasks, which can be effectively managed with strategic consideration of message-passing and supervision signal edges.

\paragraph{w/ varying extents of diversity filtering Sec~\ref{sec:diverse-filtering}} In Table~\ref{table:ablation_appendix}, we reported the results without using diversity filtering. In this section, we discuss the design choice of varying extents of diversity while filtering by comparing it with diversity filtering using static thresholds. To compare, we perform stricter filtering by choosing dissimilar samples outside of -0.5 std of the similarity distribution. On the other hand, we choose more lenient sampling by picking samples under +0.5 std of the distribution. In other words, for the stricter case we pick samples in the bottom 31$\%$ data points, for lenient sampling we pick samples in the bottom $69\%$. The results are reported in Table~\ref{table:ablation_appendix}. The degradation of performance is indicative of the utility of dynamic diversity filtering.

\paragraph{replacing w/MPNet (Sec~\ref{sec:txn-bert}): }Replacing \bert with a larger model~\cite{song2020mpnet} finetuned similar to \bert, we note that while overall performance improves, the model cannot generalize to unseen samples. This could be due to MPNet's stronger capabilities in learning contextual dependencies, which may not translate to OOD data.

\end{document}